\documentclass[12]{article}

\usepackage{graphicx,amsmath,amssymb,feynmf,dsfont,latexsym,braket,pictexwd}
\usepackage{undertilde}
\usepackage[usenames,dvipsnames]{color}
\definecolor{blue}{rgb}{0,0,.7}
\definecolor{red}{rgb}{.7,0,0}
\definecolor{orange}{rgb}{1,.6,0}
\definecolor{purple}{rgb}{.4,0,.5}
\definecolor{brown}{rgb}{.4,.2,.1}
\definecolor{green}{rgb}{0,.57,0}
\usepackage[margin=2cm]{geometry} 
 
\newcommand{\bi}{\begin{itemize}}
\newcommand{\ei}{\end{itemize}}

\newcommand{\ben}{\begin{enumerate}}
\newcommand{\een}{\end{enumerate}}

\newcommand{\sL}{\mathcal{L}}

\def\dot{\!\cdot\!}

\def\bea{\begin{eqnarray}}
\def\eea{\end{eqnarray}}

\def\d{\dagger}

\def\e{\epsilon}

\newcommand{\bx}{\mathop{\raise-.8pt\hbox{\large$\Box$}}\nolimits}
\newcommand{\vnabla}{\vec\nabla}

\usepackage[normalem]{ulem}
\usepackage{verbatim}
\usepackage{makeidx}
\makeindex

\renewcommand{\author}[1]{\vspace{2ex}{\large\begin{center}
 \setlength{\baselineskip}{3ex}#1\par\end{center}}}
\renewcommand{\thanks}[1]{\footnote{#1}}

\usepackage{bbm}

\begin{document}

\centerline{\sc \large  \\ The Photon Propagator in Light-Shell Gauge}

\author{
 Howard~Georgi,\thanks{\noindent \tt georgi@physics.harvard.edu}
Greg~Kestin,\thanks{\noindent \tt kestin@physics.harvard.edu} 
Aqil~Sajjad,\thanks{\noindent \tt sajjad@physics.harvard.edu} 
 \\ \medskip
Center for the Fundamental Laws of Nature\\
Jefferson Physical Laboratory \\
Harvard University \\
Cambridge, MA 02138
 }
\begin{abstract}
We derive the photon propagator in light-shell gauge (LSG), 
introduced in \cite{Georgi:2010nq} in the context of light-shell effective
theory. 
\end{abstract} 
\newpage
\section{Introduction}

In this paper we calculate the photon propagator in what we have called
light-shell gauge (LSG). The motivation for working in LSG is described in
depth in \cite{Georgi:2010nq} (and \cite{LSETintro}) where we discuss light-shell effective theory
(LSET). We hope that LSET, for which LSG is an essential ingredient, may
eventually provide another way of looking at high-energy scattering in
gauge theories.

While perturbative computations in gauge theories are  most commonly
carried out in covariant gauges, where the procedure has been well
established \cite{QCDprac} there \textit{are} venues in which non-covariant
gauges may be preferable. In this paper we derive the photon propagator in
light-shell gauge, which is defined by the condition 
\begin{equation}
v_\mu A^\mu = 0
\label{LSG-condition1}
\end{equation}
where
\begin{equation}
v^\mu = \left(1, \hat r\right)^\mu
\label{vmu}
\end{equation}
So, in terms of the scalar potential $A^0$ and the components of $\vec A$,
(\ref{LSG-condition1}) can be written as 
\begin{equation}
A^0 = A_r
\label{LSG-condition2}
\end{equation}
where 
\begin{equation}
A_r \equiv  \hat r\dot \vec A
\end{equation}
is the radial component of $\vec A$ (and is
not to be confused with $A_r$ in the covariant tensor form). Note also that
because (\ref{vmu}) is not well-defined at the position space origin, many
of our subsequent manipulations are ill-defined there, and we expect our
propagator to make sense only in the punctured space from which the origin
is excluded.  

A gauge that shares some characteristics with LSG is radial (Fock-Schwinger) gauge \cite{Radrules} which is defined by the condition
\begin{equation}
x_\mu A^\mu = 0,
\end{equation}
and has found widespread use in QCD sum-rules \cite{QCDsum}. Shared
characteristics between LSG and radial gauge include breaking translational
invariance by choosing an origin and \textbf{coordinate dependent
  gauge condition}. As a result, it is often convenient to use a position
space formulation rather than momentum space formulation. While these
gauges share some characteristics, only LSG guarantees zero field strength
off of the light-shell \cite{Georgi:2010nq} and allows for simplification
of calculations in LSET \cite{LSETintro}. Another important difference is
that the radial gauge condition is invariant under homogeneous Lorentz
transformations, while LSG is only invariant under 
rotations about the origin.

Since we are at such an early stage (the first, as far as we know) in
exploring this gauge, we restrict our analysis to QED where we can avoid
complications that come with non-abelian theories.\footnote{We hope to
  extend this work to QCD and in the process describe attributes avoided
  herein  (e.g. ghosts).} Even in QED, we cannot use standard techniques
for calculating propagators in non-covariant gauges, such as LSG. We
therefore, along the road to the LSG propagator, present a different
derivation which we hope may prove useful in other gauges as well. 

The basic outline of our derivation is as follows. We begin by writing the
photon lagrangian in LSG in a matrix form, treating $\vec A$ and $\hat r$ as column
vectors. In particular,  we show that the
photon's kinetic energy can be written 
\begin{equation}
\sL = -\frac{1}{2}\,
\begin{pmatrix}
A_r&\vec A_\perp^T \cr
\end{pmatrix}\,M\,\begin{pmatrix}
A_r\cr\vec A_\perp \cr
\end{pmatrix}
\label{formM}
\end{equation}
where we treat $\vec A$ as a column vector and write
\begin{equation}
\vec A_\perp=\vec A-\hat r\,\hat r^T\vec A
=\vec A-\left(\hat r\cdot\vec A\right)\,\hat r
\label{aperp}
\end{equation}
Then in the
following sections we will show how from $M$ we are able to construct the
LSG propagator. 
This is not simply a matter of inverting $M$ because $\vec A_\perp$ does
not have a radial component. What we therefore need to compute is the
inverse of $M$ restricted to the subspace from which we have projected out
this (non-existent) radial component. We will see that doing so turns out
to be non-trivial since $M$ does not commute with the projection operator
in the radial direction. As a result, we cannot express $M$ in a diagonal
basis and simply take the inverse on the relevant subspace to obtain the
propagator. We therefore need to follow a slightly more involved
procedure. Our technique, we hope, may also be applicable to other
non-covariant gauges. 

\section{The Lagrangian in LSG}

We will now find the matrix $M$ in equation (\ref{formM}) starting with the standard form of the photon kinetic energy:
\begin{equation}
\sL = -\frac{1}{4}F_{\mu\nu} F^{\mu\nu} 
= \frac{1}{2}\left(\vnabla A^0 + \partial_t \vec A\right)^2
-\frac{1}{2}\left(\vnabla\times\vec A\right)^2
 \end{equation}
We then insert the LSG condition $A^0 = A_r$, giving
\begin{equation}
\sL = \frac{1}{2}
\left(\vec\nabla A_r +\partial_t \vec A\right)^2
\,-\, \frac{1}{2}\left(\vec\nabla\times\vec A\right)^2
\end{equation}
To arrive at the form given in (\ref{formM}), we manipulate the above terms one at a time. The first term can be written as
\begin{equation}
\left(\partial_t \vec A +\vnabla A_r\right)^2
= \left(\hat r \left(\partial_t +\hat r\dot\vnabla\right) A_r
+\vnabla_\perp A_r
+\partial_t \vec A_\perp\right)^2
\label{Lterm1}
\end{equation}
where (not yet in a matrix notation)
\begin{equation}
\vnabla_\perp 
= \vnabla - \hat r \left(\hat r\cdot \vnabla\right)
\label{vecnablaperp}
\end{equation}

We expand (\ref{Lterm1}) to get
 \begin{equation}
= \left(\left(\partial_t +\hat r\dot\vnabla\right) A_r\right)^2
+ \left(\partial_t \vec A_\perp\right)^2
+\left(\vnabla_\perp A_r\right)^2
+ 2\left(\vnabla_\perp A_r\right) \dot \partial_t \vec A_\perp
\end{equation}
Integrating this by parts gives
{\renewcommand{\arraystretch}{1.8}
\begin{equation}
\begin{array}
{c}
= - A_r (\partial_t +\vnabla\cdot\hat r) (\partial_t +\hat r\cdot\vnabla) A_r
- A_r \vnabla^2 A_r
+ A_r \left(\vnabla\dot\hat r\right)\left(\hat r \dot\vnabla\right) A_r
\\-\vec A_\perp \cdot \partial_t^2 \vec A_\perp
- A_r \partial_t \vnabla_\perp \cdot A_\perp
- \vec A_\perp \cdot \vnabla_\perp \partial_t A_r 
\end{array}
\label{Lagrangian-piece1}
\end{equation}
}
For the $\left(\vnabla\times \vec A\right)^2$ term we can write
\begin{equation}
\left(\vnabla \times \vec A\right)^2
= \left(\vnabla \times A_r \hat r
+\vnabla \times \vec A_\perp\right)^2
\label{Lagrangian-piece2-full}
\end{equation}
We can work out the $rr$, $r\perp$, $\perp r$ and $\perp\perp$ terms in this
separately by writing all the cross products explicitly in terms of
Cartesian indices and simplifying. The $rr$ term is 
\begin{equation}
\left(\vnabla \times A_r \hat r\right)^2 
= \left(\hat r\times\vnabla A_r\right)^2 
= \left(\hat r_j \nabla_k A_r\right)\left(\hat r_j \nabla_k A_r\right)
- \left(\hat r_j \nabla_k A_r\right) \left(\hat r_k \nabla_j A_r\right)
\end{equation}
\begin{equation}
= \left(\vnabla A_r\right)^2 
- \left(\hat r_k \nabla_k A_r\right) \left(\hat r_j \nabla_j A_r\right)
\end{equation}
\begin{equation}
= \left(\vnabla A_r\right)^2 
- \left(\hat r \dot\vnabla A_r\right)^2
\end{equation}
Integrating this by parts gives
\begin{equation}
= -A_r \nabla^2 A_r 
+ A_r \left(\vnabla\dot\hat r\right)\left(\hat r \dot\vnabla\right) A_r
\label{Lagrangian-piece2a}
\end{equation}
The $\perp\perp$ term is
\begin{equation}
\left(\vnabla\times\vec A_\perp\right)\dot\left(\vnabla\times \vec A_\perp\right)
= \left(\nabla_j A_\perp^k\right) \left(\nabla_j A_\perp^k\right)
- \left(\nabla_j A_\perp^k\right) \left(\nabla_k A_\perp^j\right)
\end{equation}
\begin{equation}
= -\vec A_\perp \dot \nabla^2 \vec A_\perp
+ \left(\vec A_\perp \dot \vnabla\right) \left(\vnabla \dot \vec A_\perp\right)
\label{Lagrangian-piece2b}
\end{equation}
Similarly, it can be shown that the $r\perp$ and $\perp r$ terms are
\begin{equation}
\left(\vnabla \times A_r \hat r\right) \dot \left(\vnabla \times \vec A_\perp\right)
= A_r \left(\vnabla\dot \hat r\right) \left(\vnabla \dot \vec A_\perp\right)
\label{Lagrangian-piece2c}
\end{equation}
and
\begin{equation}
\left(\vnabla \times \vec A_\perp\right) \dot \left(\vnabla \times A_r \hat r\right)
= \left(\vec A_\perp \dot \vnabla\right) \left(\hat r\dot\vnabla A_r\right)
\label{Lagrangian-piece2d}
\end{equation}

Combining all the terms from (\ref{Lagrangian-piece1}), (\ref{Lagrangian-piece2a}), (\ref{Lagrangian-piece2b}), (\ref{Lagrangian-piece2c}),  and (\ref{Lagrangian-piece2d}), we can write the Lagrangian in the matrix form in (\ref{formM}) repeated below
\begin{equation}
\sL = -\frac{1}{2}\,
\begin{pmatrix}
A_r&\vec A_\perp^T \cr
\end{pmatrix}\,M\,\begin{pmatrix}
A_r\cr\vec A_\perp \cr
\end{pmatrix}
\label{Matrix-form}
\end{equation} 
where we now know the matrix $M$ is given by
\begin{equation}
M = \begin{pmatrix}
(\partial_t +\vnabla \dot\hat r)\cr
\vec\nabla    \cr
  \end{pmatrix}
  \begin{pmatrix}
(\partial_t+\hat r\dot \vnabla)&\vec\nabla^T    \cr
  \end{pmatrix}  
  +\begin{pmatrix}
0&0    \cr
0&  I\,\bx\,\cr\end{pmatrix}
\label{matrix-lightshell-gauge}
\end{equation}

Now things get a little complicated.  The $4\times4$ matrix differential
operator $M$ is invertible, but its inverse is not the propagator we want.
The LSG propagator is the inverse of $M$ restricted to the subspace from
which we have projected out the (non-existent) radial component of $\vec
A_\perp$. Let $P$ be the projection operator onto the radial direction of
$\vec A$.
Then the inverse we are looking for is the operator $D$
satisfying 
\begin{equation}
\begin{array}{c}
\displaystyle P\,D=D\,P=0
\\ \displaystyle
\left(I-P\right)\,M\,\left(I-P\right)\,D
=D\,\left(I-P\right)\,M\,\left(I-P\right)
=\left(I-P\right)
\end{array}
\label{D-subspace-inverse-property}
\end{equation}
Because $P$ does not commute with $M$, we cannot simply invert $M$ and then
project onto the relevant subspace. Instead, we will use
a 2-step procedure.   We will first show how the
linear algebra of this 2-step procedure works in general, and then apply it
to the LSG propagator in particular. 

 \section{\label{inversion}Inversion on a subspace}

Our aim is to take an invertible matrix $M$, and find its inverse restricted to the
subspace projected onto by $\left(I-P\right)$, where $P$ is a projection
operator onto a subspace and $I$ is the identity matrix. That is, we wish
to find the matrix $D$ satisfying (\ref{D-subspace-inverse-property}). 
There are two steps.  Step one (which, for LSG, we will put off until later and
relegate to an appendix)
is to find the inverse of $M^{-1}$ on the space projected onto by
$P$.  That is, we find an operator $\nu$ satisfying
\begin{equation}
\nu\,P=P\,\nu=\nu
\quad\quad\quad
\nu\,P\,M^{-1}\,P=P\,M^{-1}\,P\,\nu=P
\label{nu-condition}
\end{equation}
Then in step two we consider the following operator:
\begin{equation}
D = M^{-1} 
- M^{-1} \,\nu\, M^{-1}
= M^{-1} 
- M^{-1} \,P\,\nu\,P\, M^{-1},
\label{B}
\end{equation}
It is straightforward to apply (\ref{nu-condition}) to see that $D$
satisfies (\ref{D-subspace-inverse-property}), and thus it
is the desired inversion of $M$ on the subspace projected by
$\left(I-P\right)$.

\section{Returning to the LS gauge propagator}

We now show how we can apply (\ref{B}) to find the LSG propagator. In
this and the following sections we will 
use an operator notation (discussed in more detail 
in appendix \ref{operator-notation}) 
in which differential operators, their inverses, 
and ordinary functions of coordinates are
all treated as linear operators acting on the tensor product space of our
$4$-component index space and the space of functions of the coordinates.

In this language, the 
projection operator $P$ is
\begin{equation}
P=\begin{pmatrix}0&0\cr0&\hat R\hat R^T\end{pmatrix}
\end{equation}
Since the formula (\ref{B}) for the inverse on a subspace involves the
inverse of $M$ on the full space, we must begin by finding $M^{-1}$. For
this purpose, it is convenient to note that $M$ can be written in terms of
a diagonal matrix $M_d$ and a triangular matrix $T$ as (where $I_n$ is the
$n\times n$ identity operator) 
\begin{equation}
M = TM_d T^\d,
\end{equation} 
where
\begin{equation}
M_d = \begin{pmatrix}
\left(\partial_t+\vnabla^T\hat R\right) \left(\partial_t +\hat R^T\vec\nabla\right) & 0 \cr 
0&\bx    \cr
  \end{pmatrix},
\label{D-dagger}
\end{equation}
\begin{equation}
T=  \begin{pmatrix}
1&0\cr
\vnabla\left(\partial_t+\vec\nabla^T\hat R\right)^{-1}&I_{3}    \cr
  \end{pmatrix}
\label{T}
\end{equation}
and
\begin{equation}
T^\d =  \begin{pmatrix}
1&\left(\partial_t+\hat R^T\vec\nabla\right)^{-1}\vec\nabla^T\cr 0&I_{3}    \cr
  \end{pmatrix}
\label{T-dagger}
\end{equation}
This makes inverting $M$ straightforward, and we get for $M^{-1}$
\begin{equation}
\begin{pmatrix}
\left(\partial_t +\hat R^T\vnabla\right)^{-1} \left(1 + \vnabla^T \bx^{-1} \vnabla\right) 
\left(\partial_t +\vnabla^T\hat R\right)^{-1}
& -\left(\partial_t+\hat R^T\vnabla\right)^{-1}\vnabla^T \bx^{-1} 
\cr
-\bx^{-1} \vnabla\left(\partial_t+\vec\nabla^T\hat R\right)^{-1}
 & \bx^{-1}\cr\end{pmatrix}
\label{M-inverse}
\end{equation}
The next ingredient we need is the inverse of $M^{-1}$ restricted to the
subspace.  Here it is useful to avoid the matrix structure and define a
linear operator $\mu$, as
\begin{equation}
\mu=\begin{pmatrix}0&\hat R^T\end{pmatrix}\, M^{-1}\,\begin{pmatrix}0\cr\hat R\end{pmatrix}
\label{def-mu}
\end{equation}
whence $\nu$ in (\ref{nu-condition}) is given by 
\begin{equation}
\nu=\begin{pmatrix}0&0\cr 0&\hat R\,\mu^{-1}\,\hat R^T\end{pmatrix}
\end{equation}

Now we can just use (\ref{B}) and put the pieces together to formally compute the LSG propagator. Doing so and simplifying gives the following results:
\begin{equation}
D_{rr}=\left(\partial_t+\hat R^T\vec\nabla\right)^{-1}
\,\left(1+\vec\nabla\,^T\,C\,\vec\nabla\,\right)\,
\left(\partial_t+\vec\nabla^T\hat R\right)^{-1}
\label{radial-propagator}
\end{equation}
\begin{equation}
D_{r\perp}=-\left(\partial_t+\hat R^T\vec\nabla\right)^{-1}
\vec\nabla\,^T\,C\
\label{r-perp-propagator}
\end{equation}
\begin{equation}
D_{\perp r}=
-C\,\vec\nabla\, \left(\partial_t+\vec\nabla^T\hat R\right)^{-1}
\end{equation}
\begin{equation}
D_{\perp\perp}=C
\label{prop-perp}
\end{equation}
where $C$ is given by 
\begin{equation}
C = \bx^{-1} -\bx^{-1} \,\hat R\, \mu^{-1}\, \hat R^T \,\bx^{-1}  
\label{C-def}
\end{equation}
Note that from this form we can see that $C$ is transverse; that is, if we
act with the projection operator for the transverse subspace on either side
of $C$, we get $C$. What remains to be done is to derive  an explicit form
for $C$, which is done in detail in appendix \ref{Cappendix}, with the
result 
\begin{equation}
C=-R\,\vec\nabla_\perp\,\bx^{-1}\,L^{-2}\,R\,\vec\nabla_\perp^T
+\vec L\,\bx^{-1}\,L^{-2}\,\vec L^T
\label{C}
\end{equation}
where
\begin{equation}
R\equiv |\vec R|.
\end{equation}
Since this involves $L^{-2}$, we must show that this is well defined.  
We show in appendix
\ref{L20} that because of the operators that appear on either side of
$L^{-2}$ in (\ref{C}), the $L^{-2}$ operator never acts on an $L=0$ state,
and the expression (\ref{C}) makes sense.  

Putting (\ref{C}) into (\ref{radial-propagator}-\ref{prop-perp}) gives
\begin{equation}
D_{rr}
=\left(\partial_t+\hat R^T\vec\nabla\right)^{-1}
\,\left(1-R^{-1}\,L^2\,\bx^{-1}\,R^{-1}\right)\,
\left(\partial_t+\vec\nabla^T\hat R\right)^{-1}
\label{2radial-propagator}
\end{equation}
\begin{equation}
D_{r\perp}
=\left(\partial_t+\hat R^T\vec\nabla\right)^{-1}
\,R^{-1}\,\bx^{-1}\,\vec\nabla_\perp^T\,R
\label{2r-perp-propagator}
\end{equation}
\begin{equation}
D_{\perp r}=
R\,\vec\nabla_\perp\,\bx^{-1}\,R^{-1}
\, \left(\partial_t+\vec\nabla^T\hat R\right)^{-1}
\end{equation}
\begin{equation}
D_{\perp\perp}=
-R\,\vec\nabla_\perp\,\bx^{-1}\,L^{-2}\,R\,\vec\nabla_\perp^T
+\vec L\,\bx^{-1}\,L^{-2}\,\vec L^T
\label{2prop-perp}
\end{equation}

We can also combine these into a $3\times3$ matrix form, call it $D_3$, appropriate for
unconstrained $\vec A$ fields:
{\renewcommand{\arraystretch}{2}
\begin{equation}
\begin{array}{c}
\displaystyle
\hat R\,\left(\partial_t+\hat R^T\vec\nabla\right)^{-1}
\left(\partial_t+\vec\nabla^T\hat R\right)^{-1}\,\hat R^T
+\vec L\,\bx^{-1}\,L^{-2}\,\vec L^T
\\ \displaystyle
  -\Bigl(R\,\vec\nabla_\perp\,L^{-2}
-\hat R\,\left(\partial_t+\hat R^T\vec\nabla\right)^{-1}
\,R^{-1}\Bigr)
\,L^2\,\bx^{-1}\,
\Bigl(L^{-2}\,R\,\vec\nabla_\perp^T-R^{-1}
\, \left(\partial_t+\vec\nabla^T\hat R\right)^{-1}\,\hat R^T\Bigr)
\end{array}
\label{unconstrained}
\end{equation}
}

\section{Conclusion}

Here we have derived the photon propagator in light-shell gauge. In the
process of this derivation, we have presented  a technique  that may also
be useful for calculations in other non-covariant gauges (and, we
hope, other applications). LSG is a crucial part of the construction of the
light-shell effective theory \cite{Georgi:2010nq}, which we hope may provide a
new viewpoint for high-energy scattering in gauge theories.  We also hope
that further 
insight can be gained once this method is extended to non-abelian gauge
theories.  

\section*{Acknowledgements}

We have benefited greatly from suggestions by Matthew Schwartz, Benjamin
Grinstein, Randall Kelley, Aneesh Manohar and 
David Simmons-Duffin. Some of the initial work on this project by HG was
done at the Aspen Center for Physics.
He is grateful for the support of the Center
 and National Science Foundation grant \#1066293.
This research has been supported at Harvard 
in part by the National Science Foundation
under grants PHY-0804450 and
and PHY-1067976. 

\bibliography{shell}

\appendix

\section{Operator Notation}
\label{operator-notation}
Throughout we have used a notation that involves local and non-local operators. For example, when a local operator, such as $R^{-1}$ appears, it is
\begin{equation}
R^{-1}\left(x_{1},x_{2}\right)=\frac{1}{r_{1}}\delta\left(x_{1}-x_{2}\right)
\end{equation}
and when not written, the delta function and integrations over the arguments are implicit. We also come across the operators $\bx^{-1}$, $\left(\partial_t +\hat R\dot \vec\nabla\right)^{-1}$ and $\left(\partial_t +\vec\nabla\dot\hat R\right)^{-1}$. We know that $\bx^{-1}$ is the position space propagator for a massless scalar and is given by 
\begin{equation}
\bx^{-1}(x-y) 
= -\frac{i}{4\pi^2} \frac{1}{(x-y)^2 -i\e}
 \end{equation}
To find the expression for $(\partial_t + \hat R \dot \nabla)^{-1}$, we can consider the theory with the Lagrangian
\begin{equation}
\sL =i \phi^* (\partial_t + \hat r \dot \nabla) \phi
\end{equation}
Using canonical quantization to find the 2-point function for this theory gives 
\begin{equation}
(\partial_t + \hat R \dot \vec\nabla)^{-1}
= \frac{1}{{r'}^2} \theta(t-t')
\delta(t-r -t'+r') \delta(z-z') \delta(\phi-\phi')
\label{good-guy-inverse}
\end{equation}
Similarly
\begin{equation}
(\partial_t + \vec\nabla \dot \hat R)^{-1}
= \frac{1}{r^2} \theta(t-t')
\delta(t-r -t'+r') \delta(z-z') \delta(\phi-\phi')
\label{bad-guy-inverse}
\end{equation}
\section{Derivation of C}
\label{Cappendix}

We can  find $C$ by brute force, but here we will instead use a slicker
approach, which will take advantage of
(\ref{D-subspace-inverse-property}). 
Using the formula (\ref{matrix-lightshell-gauge}) for $M$ and our result
for the propagator (\ref{radial-propagator})-(\ref{prop-perp}), it is
straightforward to see that $\left(I_{4}-P\right) M\left(I_{4}-P\right) D$
is 
\begin{equation} \begin{pmatrix}
1 &
0\cr\vec\nabla_\perp\, \left(\partial_t+\vnabla^T\hat R\right)^{-1} -\left(I_3-P_{3}\right)\,\bx\,
\,C\,\vec\nabla\, \left(\partial_t+\vnabla^T\hat R\right)^{-1}
& \left(I_3-P_{3}\right)\,\bx\,\,C
\cr\end{pmatrix}
\end{equation}
where $P_3=\hat R\,\hat R^T$, and we have used $\left(I_3-P_{3}\right)C=C$. For $D$ to be the LSG propagator, we want the 2nd row entries of $\left(I-P\right) M\left(I-P\right)D$ to be $0$ and $I_3-P_{3}$. Both these requirements are satisfied if
\begin{equation}
\left(I_{3}-P_{3}\right)\,\bx \,C = I_{3}-P_{3}
\label{ccond}
\end{equation}
 We will now use this condition to find an explicit expression for $C$. Our approach will involve first finding a basis for the space perpendicular to $\hat R$, and then acting on (\ref{ccond}) with various operators to find the components  of $C$ in this basis. We begin by identifying the proper basis. Notice that
\begin{equation}
 R\,\vec\nabla_\perp^T=i\,\left(\vec L\times\hat R\right)^T
\label{rdperp}
\end{equation}
So  $\vec L$ and $R\,\vec\nabla_\perp$ are both orthogonal to $\hat R$ and orthogonal to one another, therefore forming our basis. We can  express $\left(I_3-P_{3}\right)$ in terms of them. First note that from (\ref{rdperp}) it follows that
\begin{equation}
R\,\vec\nabla_\perp^T\,\vec\nabla_\perp\,R=-L^2
\end{equation}
 so with proper normalization we have
\begin{equation}
\left(I_3-P_{3}\right)=\vec L\,L^{-2}\,\vec L^T
-\vec\nabla_\perp\,R\,L^{-2}\,R\,\vec\nabla_\perp^T
\label{imp}
\end{equation}
Now we want to find the components of $C$. The first, and easiest component to find is computed by acting on (\ref{ccond}) with  $\vec L$ on both sides to give
\begin{equation}
\vec L^T\left(I_3-P_{3}\right)\,\bx\,C\vec L=\vec L^T\left(I_3-P_{3}\right)\vec L
\end{equation}
This is easy because $\vec L$ commutes with $\bx$, so we get
\begin{equation}
\vec L^T\,C\,\vec L=\bx^{-1}\,L^2
\label{ll}
\end{equation}
Acting on the left of (\ref{ccond}) with $\vec L^T$ and on the right with
$\vec\nabla_\perp\,R$  as follows
\begin{equation}
\vec L^T\left(I_3-P_{3}\right)\,\bx\,C\,\vec\nabla_\perp\,R=\vec L^T\left(I_3-P_{3}\right)\vec\nabla_\perp\,R
\end{equation}
works similarly once we observe $\vec L^T\,\vec\nabla_\perp\,R=0$, giving
\begin{equation}
\vec L^T\,C\,\vec\nabla_\perp\,R=0
\label{ld}
\end{equation}
The final two matrix elements require the commutator
\begin{equation}
\Bigl[R\,\vec\nabla_\perp^T\,,\,\bx\,\Bigr]
=2R^{-2}\,L^2\,\hat R^T
\label{crucom}
\end{equation}
We now take a detour to demonstrate this commutator relation.
We can write
\begin{equation}
\bx=\partial_t^2-\left(\vec\nabla^T\hat R\right)\,\left(\hat R^T\vec\nabla\right)
+L^2\,R^{-2}
\label{bx}
\end{equation}
The middle term in (\ref{bx}) can be written
\begin{equation}
\begin{array}{c}
\left(\vec\nabla^T\hat R\right)\,\left(\hat R^T\vec\nabla\right)
=\left(\vec\nabla^T\vec R\right)\,R^{-2}\,\left(\vec R^T\vec\nabla\right)
\\
=\left(\vec R^T\vec\nabla\right)\,R^{-2}\,\left(\vec R^T\vec\nabla\right)
+3R^{-2}\,\left(\vec R^T\vec\nabla\right)
=R^{-2}\,\left(\left(\vec R^T\vec\nabla\right)^2
+\left(\vec R^T\vec\nabla\right)\right)
\end{array}
\end{equation}
We chose this particular form  because 
$\left(\vec R^T\vec\nabla\right)$ is a scaling
operator that  counts the total powers $R$ or $1/\vec\nabla$.\footnote{Note
also that the last form is trivial to remember because it vanishes for
$r^a$ with $a=0$ or $-1$ as it should.}  So this term commutes with
$R\,\vec\nabla_\perp^T$ and the only term in $\bx$ that fails to commute is $L^2\,R^{-2}$. 

The factors of $R$ commute with both $\vec\nabla_\perp^T$ and $L^2$,
so we just need to consider
\begin{equation}
\Bigl[R\,\vec\nabla_\perp^T\,,\,L^2\,\Bigr]
\end{equation}
Using (\ref{rdperp}),
we can write this in components, as
\begin{equation}
\Bigl[\,i\,\epsilon_{abc}L_b\hat R_c\,,\,L_dL_d\,\Bigr]
\end{equation}
\begin{equation}
=i\,\epsilon_{abc}L_b\,\left(
L_d\,\Bigl[\,\hat R_c\,,\,L_d\,\Bigr]
+\Bigl[\,\hat R_c\,,\,L_d\,\Bigr]\,L_d
\right)
\end{equation}
\begin{equation}
=-\epsilon_{abc}\epsilon_{cde}L_b\,\left(
L_d\hat R_e+\hat R_eL_d
\right)
\end{equation}
\begin{equation}
=-L_b\,\left(\bigl[L_a,\hat R_b\bigr]+2\hat R_bL_a-2L_b\hat R_a
-\bigl[\hat R_a,L_b\bigr]\right)
\label{last}
\end{equation}
The first and fourth terms in (\ref{last}) cancel each another.  The
second term vanishes because $\vec L\cdot\hat R=0$.
The third term gives
\begin{equation}
\Bigl[R\,\vec\nabla_\perp^T\,,\,L^2\,\Bigr]
=2L^2\,\hat R^T
\end{equation}
or
\begin{equation}
\Bigl[R\,\vec\nabla_\perp^T\,,\,\bx\,\Bigr]
=2R^{-2}\,L^2\,\hat R^T
\label{provedcrucom}
\end{equation}
which is (\ref{crucom}).

We  now return to the derivation of $C$, but note that (\ref{provedcrucom})
vanishes when acting on $C$. So, acting with $R\vec\nabla_\perp^T$ on the
left and $\nabla_\perp\,R$ on the right gives
\begin{equation} 
R\,\vec\nabla_\perp^T\,\bx\,C\,\vec\nabla_\perp\,R=
\bx\,R\,\vec\nabla_\perp^T\,C\,\vec\nabla_\perp\,R=-L^2
\label{delcal2}
\end{equation}
implying
\begin{equation}
R\,\vec\nabla_\perp^T\,C\,\vec\nabla_\perp R=-\bx^{-1}\,L^2
\label{dd}
\end{equation}
In the same way we can see that the last component is zero
\begin{equation}
R\,\vec\nabla_\perp^T\,\bx\,C\,\vec L=
\bx\,R\,\vec\nabla_\perp^T\,C\,\vec L=0
\label{dl}
\end{equation}
Combining (\ref{ll}), (\ref{ld}), (\ref{dd}) and (\ref{dl}) with (\ref{imp})
gives
\begin{equation}
C=-R\,\vec\nabla_\perp\,\bx^{-1}\,L^{-2}\,R\,\vec\nabla_\perp^T
+\vec L\,\bx^{-1}\,L^{-2}\,\vec L^T
\end{equation}

\section{Does $L^{-2}$ make sense?}
\label{L20}
The derivation
of $C$ (in appendix \ref{Cappendix}) formally involves the inverse of
$L^2$, and of course this makes no sense on $L=0$ states.  But all we
actually need is for (\ref{imp}) to make sense acting on arbitrary
functions, so that
\begin{equation}  
\left(I_3-P_{3}\right)\vec f(\vec r)=\vec L\,L^{-2}\,\vec L^T
\vec f(\vec r)-\vec\nabla\,R\,L^{-2}\,R\,\vec\nabla_\perp^T\vec f(\vec r)
\label{imp2}
\end{equation}
This is perfectly well-defined, because
if either the $\vec L^T\vec f(\vec r)$ or $R\,\vec\nabla_\perp^T\vec
f(\vec r)$ component has zero angular momentum, then that component itself
is zero. 
This can been seen by first noting that if $L^2$ acting on either of these
components  is zero, then the component must be a function of the radius
only, call it $g(r)$. If we  integrate $g(r)$ over  $d\Omega$, we get
$4\pi g(r)$, but at the same time  we see that integrating either component
over $d\Omega$ must be zero because in both cases 
we are integrating a total derivative over
a closed surface. Therefore $g(r)$, which denotes either component, is
necessarily zero.

\end{document}